\documentclass{article}
\usepackage[english]{babel}
\usepackage[cp1251]{inputenc}
\begin{document}
\begin{center}
\textbf{\Large{Some decomposition formulas of generalized
hypergeometric functions and formulas of an analytic continuation
of the Clausen function}}\\
\medskip
\textbf{A. Hasanov}\\
\medskip
Institute of Mathematics, Uzbek Academy of Sciences,\\ 29, F.
Hodjaev street, Tashkent 100125, Uzbekistan\\
E-mail: anvarhasanov@yahoo.com
\end{center}
\medskip
\begin{abstract}

In this paper, using similar symbolical method of Burchnall and
Chaundy formulas of expansion for the generalized hypergeometric
function were constructed. By means of the found formulas of
expansion the formulas of an analytic continuation for
hypergeometric function of Clausen is defined. The obtained
formulas of an analytic continuation express known hypergeometric
Appell function $ F_2 \left( {a;b_1 ,b_2 ;c_1 ,c_2 ;x,y} \right) $ which theory is well studied.

\end{abstract}

\section{Introduction and definitions}

Hardly there is a necessity to speak about importance of
properties of hypergeometric functions for any scientist and the
engineer dealing with practical application of differential
equations. The solution of various problems concerning a thermal
conduction and dynamics, electromagnetic oscillations and
aerodynamics, a quantum mechanics and the theory of potentials,
leads to hypergeometric functions. More often they appear at
solving of partial differential equations by the method of a
separation of variables. A variety of the problems leading
hypergeometric functions, has called fast growth of number of the
functions, applied in applications (for example, in the monography
[3] 205 hypergeometric functions are studied). There were
monographies and papers on the theory of special functions. But in
these monographs there is no formula of expansion and an analytic
continuation of the generalized hypergeometric function. In this
paper, using similar symbolical method Burchnall and Chaundy, we
shall construct formulas of expansion for the generalized
hypergeometric function. By means of the obtained formulas of
expansion we find the formulas of an analytic continuation of
hypergeometric function of Clausen. The found formulas of an
analytic continuation express known hypergeometric Appell function
([1], p. 14, (12), p.28, (2) and [2])
$$
F_2 \left( {a;b_1 ,b_2 ;c_1 ,c_2 ;x,y} \right) = \displaystyle
\sum\limits_{i,j = 0}^\infty  {} \displaystyle \frac{{\left( a
\right)_{i + j} \left( {b_1 } \right)_i \left( {b_2 } \right)_j
}}{{\left( {b_1 } \right)_i \left( {b_2 } \right)_j i!j!}}x^i y^j
,\left| x \right| + \left| y \right| < 1,
$$
$$
\begin{array}{l}
F_2 \left( {a;b_1 ,b_2 ;c_1 ,c_2 ;x,y} \right) = \displaystyle
\frac{{\Gamma \left( {c_1 } \right)\Gamma \left( {c_2 }
\right)}}{{\Gamma \left( {b_1 } \right)\Gamma
\left( {c_1  - b_1 } \right)\Gamma \left( {b_2 } \right)\Gamma \left( {c_2  - b_1 } \right)}} \times  \\
\times \displaystyle \int\limits_0^1 {\displaystyle
\int\limits_0^1 {} \xi ^{b_1 - 1} \eta ^{b_2  - 1} \left( {1 - \xi
} \right)^{c_1 - b_1  - 1} } \left( {1 - \eta } \right)^{c_2  -
b_2  - 1} \left( {1 - x\xi  - y\eta }
\right)^{ - a} d\xi d\eta ,{\mathop{\rm Re}\nolimits} c_i  > {\mathop{\rm Re}\nolimits} b_i  > 0,i = 1,2., \\
\end{array}
$$
which theory is well studied.

In the Gaussian [1, 2], hypergeometric series $ {}_2F_1 \left(
{a,b;c;x} \right)$ there are two numeration parameters $ a,b $ and
one denominator parameter $c$. A natural generalization of this
series is accomplished by introducing any arbitrary number of
numerator and denominator parameters. The resulting series [1-3]
$$
{}_pF_q \left[ {\begin{array}{*{20}c}
{\alpha _1 ,\alpha _1 ,...,\alpha _p ;}  \\
{\beta _1 ,\beta _1 ,...,\beta _q ;}  \\
\end{array}z} \right] = \displaystyle \sum\limits_{m = 0}^\infty  {} \frac{{\left( {\alpha _1 }
\right)_m \left( {\alpha _2 } \right)_m ...\left( {\alpha _p }
\right)_m }}{{\left( {\beta _1 } \right)_m \left( {\beta _2 }
\right)_m ...\left( {\beta _q } \right)_m m!}}z^m ,\eqno (1.1)
$$
where $\left( \lambda  \right)_n$ is the Pochhammer symbol defined
$$
\displaystyle \left( \lambda  \right)_n  = \frac{{\Gamma \left(
{\lambda  + n} \right)}}{{\Gamma \left( \lambda
\right)}},\,\lambda  \ne 0, - 1, - 2,...,\eqno (1.2)
$$
is known as the generalized Gauss series, or simply, the
generalized hypergeometric series. Here $ p $ and $ q $ are
positive integers or zero (interpreting an empty product as 1),
and we assume that the variable $ z $, the numerator parameters $
\alpha _1 ,...,\alpha _p $, and the denominator parameters $ \beta
_1 ,...,\beta _q $ take on complex values, provided that
$$
\displaystyle \beta _i  \ne 0,\, - 1,\, - 2,\,...;\,j =
1,\,...,\,q.\eqno (1.3)
$$
Supposing that none of the numerator parameters is zero or a
negative integer, and with the usual restriction (1.3), the series
in (1.1)

(I) converges for $ \left| z \right| < \infty $ if $ p \le q $,

(II) converges for $ \left| z \right| < \infty $ if $ p = q + 1 $,
and

(III) diverges for all $ z $ , , if $ p > q + 1 $. \\
Furthermore, if we set
$$
\omega  = \displaystyle \sum\limits_{i = 1}^q {} \beta _i  -
\sum\limits_{i = 1}^p {} \alpha _i,\eqno (1.4)
$$
it is known that the $ {}_pF_q $ series, with $ p = q + 1$, is

(IV) absolutely converges for   if $ {\mathop{\rm Re}\nolimits}
\,(\omega ) > 0 $,

(V) conditionally convergent for, $ z \ne 1 $, if $
 - 1 < \,{\mathop{\rm Re}\nolimits} \,(\omega )\, \le 0 $, and

(VI) divergent for $ \left| z \right| = 1$ if $ \,{\mathop{\rm
Re}\nolimits} \,(\omega )\, \le  - 1$.\\
The formula of an analytic continuation of hypergeometric
functions have important value in the theory hypergeometric
functions. The formula of an analytic continuation for
hypergeometric function of Gauss ${}_2F_1 \left( {a,b;c;x}
\right)$ can be found in the monography [1, 2]. In the paper ([4],
p. 704, (3)), for hypergeometric function of Clausen (1.1) in the
case of $p = 3,\,\,q = 2$ the formulas of an analytic continuation
were found by the author. One of the found formulas has a form of
$$
{}_3F_2 \left[ {\begin{array}{*{20}c}
   {a_1 ,a_2 ,a_3 }  \\
   {b_1 ,b_2 }  \\
\end{array}x} \right] = F_R \left[ {\begin{array}{*{20}c}
   {a_1 ,a_2 ,a_3 }  \\
   {b_1 ,b_2 }  \\
\end{array}x} \right] + \xi \left[ {\begin{array}{*{20}c}
   {a_1 ,a_2 ,a_3 }  \\
   {b_1 ,b_2 }  \\
\end{array}x} \right],\eqno (1.5)
$$
where
$$
\begin{array}{l}
F_R \left[ {\begin{array}{*{20}c}
{a_1 ,a_2 ,a_3 }  \\
{b_1 ,b_2 }  \\
\end{array}x} \right] = \displaystyle \frac{{\Gamma \left( {b_1 } \right)\Gamma \left( {b_2 } \right)\Gamma
\left( {b_1  + b_2  - a_1  - a_2  - a_3 } \right)}}{{\Gamma \left(
{a_1 } \right)\Gamma \left( {b_1  + b_2  - a_1  - a_2 } \right)\Gamma \left( {b_1  + b_2  - a_1  - a_3 } \right)}} \\
\displaystyle \times \sum\limits_{n = 0}^\infty  {} \frac{{\left(
{b_1  - a_1 } \right)_n \left( {b_2  - a_1 } \right)_n \left( {b_1
+ b_2  - a_1  - a_2  - a_3 } \right)_n }}{{\left( {b_1  + b_2  - a_1  - a_2 }
\right)_n \left( {b_1  + b_2  - a_1  - a_3 } \right)_n n!}} \\
\displaystyle \times {}_2F_1 \left( {a_2 ,a_3 ;a_1  + a_2  + a_3  - b_1  - b_2  - n + 1;1 - x} \right), \\
\end{array}\eqno (1.6)
$$
$$
\begin{array}{l}
\xi \left[ {\begin{array}{*{20}c}
{a_1 ,a_2 ,a_3 }  \\
{b_1 ,b_2 }  \\
\end{array}x} \right] = \displaystyle \frac{{\Gamma \left( {b_1 } \right)\Gamma \left( {b_2 } \right)\Gamma
\left( {a_1  + a_2  + a_3  - b_1  - b_2 } \right)}}{{\Gamma \left(
{a_1 } \right)\Gamma \left( {a_2 } \right)\Gamma \left( {a_3 } \right)}} \\
\displaystyle \times x^{a_1  - b_1  - b_2  + 1} \left( {1 - x}
\right)^{b_1  + b_2  - a_1  - a_2  - a_3 } \displaystyle
\sum\limits_{n = 0}^\infty  {} \displaystyle \frac{{\left( {b_1  -
a_1 } \right)_n \left( {b_2  - a_1 } \right)_n }}{{\left( {b_1  +
b_2  - a_1  - a_2  - a_3  + 1} \right)_n n!}}\left( {\displaystyle \frac{{x - 1}}{x}}\right)^n  \\
\displaystyle \times {}_2F_1 \left( {1 - a_2 ,1 - a_3 ;b_1  + b_2  - a_1  - a_2  - a_3  + n + 1;1 - x} \right). \\
\end{array} \eqno (1.7)
$$
From identity (1.5) follows that the obtained formula of an
analytic continuation for hypergeometric function of Clausen $
{}_3F_2 $ is very bulky and not convenient in applications, and
does not express through known hypergeometric functions.

\section{Preliminaries}

Burchnall and Chaundy ([5] and [6]) and (Chaundy [7])
systematically presented a number of expansion and decomposition
formulas for double hypergeometric functions in series of simpler
hypergeometric functions. Their method is based upon the following
inverse pairs of symbolic operators:
$$
\nabla _{x,y} \left( h \right): = \displaystyle \frac{{\Gamma
\left( h \right)\Gamma \left( {\delta  + \delta ' + h}
\right)}}{{\Gamma \left( {\delta  + h} \right)\Gamma \left(
{\delta ' + h} \right)}} = \displaystyle \sum\limits_{i =
0}^\infty {} \displaystyle \frac{{\left( { - \delta } \right)_i
\left( { - \delta '} \right)_i }}{{\left( h \right)_i i!}}, \eqno
(2.1)
$$
$$
\begin{array}{l}
\Delta _{x,y} \left( h \right): = \displaystyle \frac{{\Gamma
\left( {\delta  + h} \right)\Gamma \left( {\delta ' + h}
\right)}}{{\Gamma \left( h \right) \Gamma \left( {\delta  + \delta
' + h} \right)}} = \displaystyle \sum\limits_{i = 0}^\infty  {}
\frac{{\left( { - \delta } \right)_i \left( { - \delta '}
\right)_i }}{{\left( {1 - h - \delta  - \delta '} \right)_i i!}} \\
\,\,\,\,\,\,\,\,\,\,\,\,\,\,\,\,\, = \displaystyle \sum\limits_{i
= 0}^\infty  {} \displaystyle \frac{{\left( { - 1} \right)^i
\left( h \right)_{2i} \left( { - \delta } \right)_i \left( { -
\delta '} \right)_i }}{{\left( {h + i - 1} \right)_i \left(
{\delta  + h} \right)_i \left( {\delta ' + h} \right)_i i!}},\,\,
\left( {\delta : = x\frac{\partial }{{\partial x}};\,\,\delta ': = y\frac{\partial }{{\partial y}}} \right). \\
\end{array} \eqno (2.2)
$$
Indeed, as already observed by Srivastava and Karlsson ([3], pp.
332-333), the aforementioned method of Burchnall and Chaundy (cf.
[5] and [6]) was subsequently applied  by Pandey [8] and
Srivastava [9] in order to derive the corresponding expansion and
decomposition formulas for the triple hypergeometric functions $
F_A^{\left( 3 \right)} ,\,F_E ,\,F_K ,\,F_M ,\,F_P $  and $ F_T
,\,H_A ,\,H_C$, respectively (see, for definitions, [3, Section
1.5] and [10], p. 66 et seq.), by Singhal and Bhati [11], Hasanov
and Srivastava [14, 15] for deriving analogous multiple-series
expansions associated with several multivariable hypergeometric
functions. We now introduce here the following analogues of
Burchnall-Chaundy symbolic operators $ \nabla _{x,y} $ and $
\Delta _{x,y}$ defined by (2.1) and (2.2), respectively:
$$
H_{x_1 ,...,x_l } \left( {\alpha ,\beta } \right): = \displaystyle
\frac{{\Gamma \left( \beta  \right)\Gamma \left( {\alpha  + \delta
_1  +  \cdot \cdot  \cdot  + \delta _l } \right)}}{{\Gamma \left(
\alpha \right)\Gamma \left( {\beta  + \delta _1  +  \cdot  \cdot
\cdot  + \delta _l } \right)}} = \displaystyle \sum\limits_{k_1 ,
\cdot  \cdot \cdot ,k_l = 0}^\infty  {} \displaystyle
\frac{{\left( {\beta  - \alpha } \right)_{k_1  + \cdot  \cdot
\cdot  + k_l } \left( { - \delta _1 } \right)_{k_1 } \cdot  \cdot
\cdot \left( { - \delta _l } \right)_{k_l } }}{{\left( \beta
\right)_{k_1  + \cdot  \cdot \cdot  + k_l } k_1 ! \cdot  \cdot
\cdot k_l !}}\eqno (2.3)
$$
and
$$
\bar H_{x_1 ,...,x_l } \left( {\alpha ,\beta } \right): =
\displaystyle \frac{{\Gamma \left( \alpha  \right)\Gamma \left(
{\beta  + \delta _1  +  \cdot  \cdot  \cdot  + \delta _l }
\right)}}{{\Gamma \left( \beta  \right)\Gamma \left( {\alpha  +
\delta _1  +  \cdot  \cdot \cdot  + \delta _l } \right)}} =
\displaystyle \sum\limits_{k_1 ,...k_l  = 0}^\infty  {}
\displaystyle \frac{{\left( {\beta  - \alpha } \right)_{k_1  +
\cdot  \cdot \cdot  + k_l } \left( { - \delta _1 } \right)_{k_1 }
\cdot  \cdot \cdot \left( { - \delta _l } \right)_{k_l }
}}{{\left( {1 - \alpha - \delta _1  -  \cdot \cdot  \cdot  -
\delta _l } \right)_{k_1  + \cdot  \cdot  \cdot + k_l } k_1 !
\cdot  \cdot  \cdot k_l !}} \eqno (2.4)
$$
$$
\left( {\delta _j : = x_j \displaystyle \frac{\partial }{{\partial
x_j }},\,j = 1,...,l;\,\,l \in N: = \left\{ {1,2,3,...} \right\}}
\right).
$$
By means of operators (2.3) - (2.4) we shall construct functional
identities for the generalized hypergeometric function (1.1).

\section{Functional identities}

Similarly just as in the papers [5, 6], the following functional
identities
$$
\begin{array}{l}
{}_pF_q \left[ {\begin{array}{*{20}c}
{\alpha _1 ,\alpha _2 ,...,\alpha _p ;}  \\
{\beta _1 ,\beta _2 ,...,\beta _p ;}  \\
\end{array}x} \right] = H_x \left( {\alpha _p ,\beta _q }
\right)\,\,{}_{p - 1}F_{q - 1} \left[ {\begin{array}{*{20}c}
{\alpha _1 ,\alpha _2 ,...,\alpha _{p - 1} ;}  \\
{\beta _1 ,\beta _2 ,...,\beta _{q - 1} ;}  \\
\end{array}x} \right],p \ge 3,\,\,q \ge 2,
\end{array} \eqno (3.1)
$$
$$
\begin{array}{l}
{}_{p - 1}F_{q - 1} \left[ {\begin{array}{*{20}c}
{\alpha _1 ,\alpha _2 ,...,\alpha _{p - 1} ;}  \\
{\beta _1 ,\beta _2 ,...,\beta _{q - 1} ;}  \\
\end{array}x} \right] = \bar H_x \left( {\alpha _p ,\beta _q }
\right)\,{}_pF_q \left[ {\begin{array}{*{20}c}
{\alpha _1 ,\alpha _2 ,...,\alpha _p ;}  \\
{\beta _1 ,\beta _2 ,...,\beta _p ;}  \\
\end{array}x} \right],p \ge 3,\,\,q \ge 2,
\end{array} \eqno (3.2)
$$
$$
\begin{array}{l}
{}_pF_q \left[ {\begin{array}{*{20}c}
{\alpha _1 ,\alpha _2 ,...,\alpha _p ;}  \\
{\beta _1 ,\beta _2 ,...,\beta _q ;}  \\
\end{array}x} \right] = H_x \left( {\alpha _p ,\beta _q } \right)H_x \left( {\alpha _{p - 1} ,\beta _{q - 1} }
\right){}_{p - 2}F_{q - 2} \left[ {\begin{array}{*{20}c}
{\alpha _1 ,\alpha _2 ,...,\alpha _{p - 2} ;}  \\
{\beta _1 ,\beta _2 ,...,\beta _{q - 2} ;}  \\
\end{array}x} \right],
p \ge 4,\,\,q \ge 3, \\
\end{array}\eqno (3.3)
$$
$$
\begin{array}{l}
\,{}_{p - 2}F_{q - 2} \left[ {\begin{array}{*{20}c}
{\alpha _1 ,\alpha _2 ,...,\alpha _{p - 2} ;}  \\
{\beta _1 ,\beta _2 ,...,\beta _{q - 2} ;}  \\
\end{array}x} \right] = \bar H_x \left( {\alpha _p ,\beta _q }
\right)\bar H_x \left( {\alpha _{p - 1} ,\beta _{q - 1} } \right)\,{}_pF_q \left[ {\begin{array}{*{20}c}
{\alpha _1 ,\alpha _2 ,...,\alpha _p ;}  \\
{\beta _1 ,\beta _2 ,...,\beta _p ;}  \\
\end{array}x} \right], p \ge 4,\,\,q \ge 3
\end{array}\eqno (3.4)
$$
are true. Validity of functional identities (3.1) - (3.4) can be
proved by means of transformation Mellin (for example see [12]).

\section{Formulas of expansion for the generalized hypergeometric function}

Applying operators (2.3) - (2.3), from functional identities we
define the following expansions for the generalized hypergeometric
function (1.1)
$$
\begin{array}{l}
{}_pF_q \left[ {\begin{array}{*{20}c}
{\alpha _1 ,\alpha _2 ,...,\alpha _p ;}  \\
{\beta _1 ,\beta _2 ,...,\beta _p ;}  \\
\end{array}x} \right] \\
= \displaystyle \sum\limits_{i = 0}^\infty  {} \displaystyle
\frac{{\left( { - 1} \right)^i \left( {\alpha _1 } \right)_i
\left( {\alpha _2 } \right)_i ...\left( {\alpha _{p - 1} }
\right)_i \left( {\beta _q - \alpha _p } \right)_i }}{{\left(
{\beta _1 } \right)_i \left( {\beta _2 } \right)_i ...\left(
{\beta _{q - 1} } \right)_i \left( {\beta _q } \right)_i i!}}x^i
{}_{p - 1}F_{q - 1} \left[ {\begin{array}{*{20}c}{\alpha _1  + i,\alpha _2  + i,...,\alpha _{p - 1}  + i;}  \\
{\beta _1  + i,\beta _2  + i,...,\beta _{q - 1}  + i;}  \\
\end{array}x} \right], p \ge 3,\,\,q \ge 2, \\
\end{array}\eqno (4.1)
$$
$$
\begin{array}{l}
\,{}_{p - 1}F_{q - 1} \left[ {\begin{array}{*{20}c}
{\alpha _1 ,\alpha _2 ,...,\alpha _{p - 1} ;}  \\
{\beta _1 ,\beta _2 ,...,\beta _{q - 1} ;}  \\
\end{array}x} \right] \\
= \displaystyle \sum\limits_{i = 0}^\infty  {} \displaystyle
\frac{{\left( {\alpha _1 } \right)_i \left( {\alpha _2 } \right)_i
...\left( {\alpha _{p - 1} } \right)_i \left( {\beta _q  - \alpha
_p } \right)_i }}{{\left( {\beta _1 } \right)_i \left( {\beta _2 }
\right)_i ...\left( {\beta _p } \right)_i i!}}x^i \,{}_pF_q \left[
{\begin{array}{*{20}c} {\alpha _1  + i,\alpha _2  + i,...,\alpha _{p - 1}  + i,\alpha _p ;}  \\
{\beta _1  + i,\beta _2  + i,...,\beta _q  + i;}
\end{array}x} \right], p \ge 3,\,\,q \ge 2,
\end{array}\eqno (4.2)
$$
$$
\begin{array}{l}
{}_pF_q \left[ {\begin{array}{*{20}c}
{\alpha _1 ,\alpha _2 ,...,\alpha _p ;}  \\
{\beta _1 ,\beta _2 ,...,\beta _q ;}  \\
\end{array}x} \right] \\
= \displaystyle \sum\limits_{i,j = 0}^\infty  {} \displaystyle
\frac{{\left( { - 1} \right)^{i + j} \left( {\alpha _1 }
\right)_{i + j} \left( {\alpha _2 } \right)_{i + j} ...\left(
{\alpha _{p - 2} } \right)_{i + j} \left( {\alpha _{p - 1} }
\right)_i \left( {\beta _{q - 1}  - \alpha _{p - 1} } \right)_j
\left( {\beta _q  - \alpha _p } \right)_i }}{{\left( {\beta _1 }
\right)_{i + j} \left( {\beta _2 } \right)_{i + j} ...\left(
{\beta _{q - 2} } \right)_{i + j} \left( {\beta _{q - 1} }
\right)_{i + j} \left( {\beta _q } \right)_i i!j!}}x^{i + j} \,\, \\
\times {}_{p - 2}F_{q - 2} \left[ {\begin{array}{*{20}c}
{\alpha _1  + i,\alpha _2  + i,...,\alpha _{p - 2}  + i;}  \\
{\beta _1  + i,\beta _2  + i,...,\beta _{q - 2}  + i;}  \\
\end{array}x} \right],p \ge 4,\,\,q \ge 3, \\
\end{array} \eqno (4.3)
$$
$$
\begin{array}{l}
\,{}_{p - 2}F_{q - 2} \left[ {\begin{array}{*{20}c}
{\alpha _1 ,\alpha _2 ,...,\alpha _{p - 2} ;}  \\
{\beta _1 ,\beta _2 ,...,\beta _{q - 2} ;}  \\
\end{array}x} \right] \\
= \displaystyle  \sum\limits_{i,j = 0}^\infty  {} \displaystyle
 \frac{{\left( {\alpha _1 } \right)_{i + j} \left( {\alpha _2 }
\right)_{i + j} ...\left( {\alpha _{p - 2} } \right)_{i + j}
\left( {\alpha _p } \right)_j \left( {\beta _{q - 1} } \right)_i
\left( {\beta _{q - 1}  - \alpha _{p - 1} } \right)_{i + j} \left(
{\beta _q  - \alpha _p } \right)_i }}{{\left( {\beta _1 }
\right)_{i + j} \left( {\beta _2 } \right)_{i + j} ...\left(
{\beta _q } \right)_{i + j}
\left( {\beta _{q - 1}  - \alpha _{p - 1} } \right)_i i!j!}}\,x^{i + j}  \\
\times {}_pF_q \left[ {\begin{array}{*{20}c} {\alpha _1  + i +
j,\alpha _2  + i + j,...,\alpha _{p - 2}  + i + j,
\alpha _{p - 1} ,\alpha _p  + j;}  \\
{\beta _1  + i + j,\beta _2  + i + j,...,\beta _q  + i + j;}  \\
\end{array}x} \right],p \ge 4,\,\,q \ge 3, \\
\end{array}\eqno (4.4)
$$
$$
\begin{array}{l}
{}_pF_q \left[ {\begin{array}{*{20}c}
{\alpha _1 ,\alpha _2 ,...,\alpha _p ;}  \\
{\beta _1 ,\beta _2 ,...,\beta _q ;}  \\
\end{array}x + y - xy} \right] \\
= \displaystyle \sum\limits_{i = 0}^\infty  {}\displaystyle
\frac{{\left( { - 1} \right)^i \left( {\alpha _1 } \right)_i
\left( {\alpha _2 } \right)_i ...\left( {\alpha _p } \right)_i
}}{{\left( {\beta _1 } \right)_i \left( {\beta _2 } \right)_i
...\left( {\beta _q } \right)_i i!}}x^i y^i {}_pF_q \left[
{\begin{array}{*{20}c} {\alpha _1  + i,\alpha _2  + i,...,\alpha _p  + i;}  \\
{\beta _1  + i,\beta _2  + i,...,\beta _q  + i;}  \\
\end{array}x + y} \right], \\
\end{array}\eqno (4.5)
$$
$$
\begin{array}{l}
{}_pF_q \left[ {\begin{array}{*{20}c}
{\alpha _1 ,\alpha _2 ,...,\alpha _p ;}  \\
{\beta _1 ,\beta _2 ,...,\beta _q ;}  \\
\end{array}x + y} \right] \\
= \displaystyle \sum\limits_{i = 0}^\infty  {} \displaystyle
\frac{{\left( {\alpha _1 } \right)_i \left( {\alpha _2 } \right)_i
...\left( {\alpha _p } \right)_i }}{{\left( {\beta _1 } \right)_i
\left( {\beta _2 } \right)_i ...\left( {\beta _q } \right)_i
i!}}x^i y^i {}_pF_q \left[ {\begin{array}{*{20}c}{\alpha _1  + i,\alpha _2  + i,...,\alpha _p  + i;}  \\
{\beta _1  + i,\beta _2  + i,...,\beta _q  + i;}  \\
\end{array}x + y - xy} \right]. \\
\end{array}\eqno (4.6)
$$
Let prove the expansion (4.1) for the generalized hypergeometric
function. From functional identity (3.1) considering an operator
(2.3), we define
$$
{}_pF_q \left[ {\begin{array}{*{20}c}
{\alpha _1 ,\alpha _2 ,...,\alpha _p ;}  \\
{\beta _1 ,\beta _2 ,...,\beta _p ;}  \\
\end{array}x} \right] = \sum\limits_{i = 0}^\infty  {} \frac{{\left( {\beta _q  - \alpha _p }
\right)_i \left( { - \delta } \right)_i }}{{\left( {\beta _q } \right)_i i!}}\,\,{}_{p - 1}F_{q - 1}
\left[ {\begin{array}{*{20}c}{\alpha _1 ,\alpha _2 ,...,\alpha _{p - 1} ;}  \\
{\beta _1 ,\beta _2 ,...,\beta _{q - 1} ;}  \\
\end{array}x} \right].\eqno (4.7)
$$
In the ([13], p. 93) it is proved that for any analytical function
$ f\left( z \right)$  takes place identities
$$
\left( { - \delta } \right)_i f\left( z \right) = \left( { -
\delta } \right)\left( {1 - \delta } \right) \cdot  \cdot  \cdot
\left( {i - 1 - \delta } \right)f\left( z \right) = \left( { - 1}
\right)^i z^i \frac{{d^i }}{{dz^i }}f\left( z \right).\eqno (4.8)
$$
Then on the basis of identity (4.8) and by virtue of the
differentiation formula for the generalized hypergeometric
function ([1], p. 153, (31))
$$
\frac{{d^i }}{{dx^i }}{}_pF_q \left[ {\begin{array}{*{20}c}
{\alpha _1 ,\alpha _2 ,...,\alpha _p ;}  \\
{\beta _1 ,\beta _2 ,...,\beta _p ;}  \\
\end{array}x} \right] = \frac{{\left( {\alpha _1 } \right)_i \left( {\alpha _2 }
\right)_i ...\left( {\alpha _p } \right)_i }}{{\left( {\beta _1 } \right)_i
\left( {\beta _2 } \right)_i ...\left( {\beta _p } \right)_i }}{}_pF_q \left[ {\begin{array}{*{20}c}
{\alpha _1  + i,\alpha _2  + i,...,\alpha _p  + i;}  \\
{\beta _1  + i,\beta _2  + i,...,\beta _p  + i;}  \\
\end{array}x} \right],\eqno (4.9)
$$
we find
$$
\begin{array}{l}
\left( { - \delta } \right)_i \,\,{}_{p - 1}F_{q - 1} \left[
{\begin{array}{*{20}c}
{\alpha _1 ,\alpha _2 ,...,\alpha _{p - 1} ;}  \\
{\beta _1 ,\beta _2 ,...,\beta _{q - 1} ;}  \\
\end{array}x} \right] \\
=\displaystyle x^i \frac{{\left( { - 1} \right)^i \left( {\alpha
_1 } \right)_i \left( {\alpha _2 } \right)_i ...\left( {\alpha _{p
- 1} } \right)_i }}{{\left( {\beta _1 } \right)_i \left( {\beta _2
} \right)_i ...\left( {\beta _{p - 1} } \right)_i }}{}_{p - 1}F_{q
- 1} \left[ {\begin{array}{*{20}c}{\alpha _1  + i,\alpha _2  + i,...,\alpha _{p - 1}  + i;}  \\
{\beta _1  + i,\beta _2  + i,...,\beta _{p - 1}  + i;}  \\
\end{array}x} \right]. \\
\end{array}\eqno (4.10)
$$
Substituting identities (4.10) into equalities (4.7), we get the
expansion (4.1). To prove these we begin with (4.5), (4.6) which
are disguised forms of Taylor's series. For we can write Taylor's
series in the form
$$
\begin{array}{l}
{}_pF_q \left[ {\begin{array}{*{20}c}
{\alpha _1 ,\alpha _2 ,...,\alpha _p ;}  \\
{\beta _1 ,\beta _2 ,...,\beta _q ;}  \\
\end{array}x - h} \right] = \displaystyle \sum\limits_{i = 0}^\infty  {} \displaystyle \frac{{h^i }}{{i!}}x^{ - i}
\left( { - \delta } \right)_i {}_pF_q \left[{\begin{array}{*{20}c} {\alpha _1 ,\alpha _2 ,...,\alpha _p ;}  \\
{\beta _1 ,\beta _2 ,...,\beta _q ;}  \\
\end{array}x} \right] \\
 =\displaystyle \sum\limits_{i = 0}^\infty  {} \displaystyle \frac{{\left( { - 1}
\right)^i \left( {\alpha _1 } \right)_i \left( {\alpha _2 }
\right)_i ...\left( {\alpha _p } \right)_i }}{{\left( {\beta _1 }
\right)_i \left( {\beta _2 } \right)_i ...\left( {\beta _q }
\right)_i i!}}h^i {}_pF_q \left[ {\begin{array}{*{20}c}
{\alpha _1  + i,\alpha _2  + i,...,\alpha _p  + i;}  \\
{\beta _1  + i,\beta _2  + i,...,\beta _q  + i;}  \\
\end{array}x} \right]. \\
\end{array}
$$
Replacing the arguments $x,\,h$ by (i) $ x + y$, $ xy $ and (ii) $
x + y - xy$, $ - xy $ , we get (4.5), (4.6). Validity of expansion
(4.1) - (4.4) can be proved similarly. In the specific case from
formulas of expansion (4.1) - (4.4) the following outcomes follow
$$
\begin{array}{l}
{}_pF_q \left[ {\begin{array}{*{20}c}
{\alpha _1 ,\alpha _2 ,...,\alpha _p ;}  \\
{\beta _1 ,\beta _2 ,...,\beta _q ;}  \\
\end{array}x} \right] \\
= \displaystyle \sum\limits_{j = 0}^\infty  {} \displaystyle
\frac{{\left( {\alpha _1 } \right)_j \left( {\alpha _2 } \right)_j
...\left( {\alpha _{p - 1} } \right)_j }}{{\left( {\beta _1 }
\right)_j \left( {\beta _2 } \right)_j ...\left( {\beta _{q - 1} }
\right)_j j!}}x^j {}_pF_q \left[ {\begin{array}{*{20}c} {\alpha _1  + j,\alpha _2  + j,...,
\alpha _{p - 1}  + j,\beta _q  - \alpha _p; }  \\
{\beta _1  + j,\beta _2  + j,...,\beta _q  + j,\beta _q; }  \\
\end{array} - x} \right], \\
\end{array} \eqno (4.11)
$$
$$
\displaystyle {}_pF_q \left[ {\begin{array}{*{20}c}
{\alpha _1 ,\alpha _2 ,...,\alpha _p ;}  \\
{\beta _1 ,\beta _2 ,...,\beta _q ;}  \\
\end{array}x} \right] = \displaystyle F_{q - 1,1,0}^{p - 1,1,0} \left[ {\begin{array}{*{20}c}
{\alpha _1 ,\alpha _2 ,...,\alpha _{p - 1} ;}  \\
{\beta _1 ,\beta _2 ,...,\beta _{q - 1} ;}  \\
\end{array}\begin{array}{*{20}c} {\beta _q  - \alpha _p ;}  \\
{\,\,\,\,\,\,\,\,\,\,\beta _q ;}  \\
\end{array}\begin{array}{*{20}c}
{ - ;}  \\
{ - ;}  \\
\end{array} - x,x} \right],\eqno (4.12)
$$
$$
{}_3F_2 \left[ {\begin{array}{*{20}c}
   {\alpha _1 ,\alpha _2 ,\alpha _3 }  \\
   {\beta _1 ,\beta _2 }  \\
\end{array}x} \right] = \left( {1 - x} \right)^{ - \alpha _1 } F_{1,1,0}^{1,2,1} \left[ {\begin{array}{*{20}c}
   {\alpha _1 ;}  \\
   {\beta _1 ;}  \\
\end{array}\begin{array}{*{20}c}
   {\alpha _2 ,\beta _2  - \alpha _3 ;}  \\
   {\,\,\,\,\,\,\,\,\,\,\,\,\,\,\,\,\beta _2 ;}  \\
\end{array}\begin{array}{*{20}c}
   {\beta _1  - \alpha _2 ;}  \\
   {\,\,\,\,\,\,\,\,\,\, - ;}  \\
\end{array}\frac{x}{{x - 1}},\frac{x}{{x - 1}}} \right],\eqno (4.13)
$$
where ([1], p. 150, (29))
$$
\displaystyle F_{l;i;j}^{p;q;k} \left[ {\begin{array}{*{20}c}
   {\left( {a_p } \right);}  \\
   {\left( {\alpha _l } \right);}  \\
\end{array}\begin{array}{*{20}c}
   {\left( {b_q } \right);}  \\
   {\left( {\beta _m } \right);}  \\
\end{array}\begin{array}{*{20}c}
   {\left( {c_k } \right);}  \\
   {\left( {\gamma _n } \right);}  \\
\end{array}x,y} \right] = \displaystyle \sum\limits_{r,s = 0}^\infty  {}
\displaystyle \frac{{\prod\limits_{j = 1}^p {} \left( {a_j }
\right)_{r + s} \prod\limits_{j = 1}^q {} \left( {b_j } \right)_r
\prod\limits_{j = 1}^k {} \left( {c_j } \right)_s
}}{{\prod\limits_{j = 1}^l {} \left( {\alpha _j } \right)_{r + s}
\prod\limits_{j = 1}^m {} \left( {\beta _j } \right)_r
\prod\limits_{j = 1}^n {} \left( {\gamma _j } \right)_s r!s!}}x^r
y^s .\eqno (4.14)
$$
Formulas (4.11) - (4.12) are remarkable that at particular values
of argument and parameters of Clausen function, we find the sums
from some series. For example, using outcome Saalschutz [2]
$$
{}_3F_2 \left[ {\begin{array}{*{20}c}
   {a,b, - n;}  \\
   {c,1 + a + b - c - n;}  \\
\end{array}1} \right] = \displaystyle \frac{{\left( {c - a} \right)_n \left( {c - b}
\right)_n }}{{\left( c \right)_n \left( {c - a - b} \right)_n
}},\,\,n \in N = \left\{ {1,2,...} \right\},\eqno (4.15)
$$
from (4.11) we determine value
$$
\sum\limits_{j = 0}^\infty  {} \displaystyle \frac{{\left( a
\right)_j \left( b \right)_j }}{{\left( c \right)_j j!}}{}_3F_2
\left[ {\begin{array}{*{20}c}
   {a + j,b + j,1 + a + b - c;}  \\
   {c + j,1 + a + b - c - n;}  \\
\end{array} - 1} \right] = \displaystyle \frac{{\left( {c - a} \right)_n \left( {c - b}
\right)_n }}{{\left( c \right)_n \left( {c - a - b} \right)_n
}},\,\,n \in N,\eqno (4.16)
$$
Also it is not difficult to define integral representations of
Clausen function
$$
{}_3F_2 \left[ {\begin{array}{*{20}c}
{\alpha _1 ,\alpha _2 ,\alpha _3 ;}  \\
{\beta _1 ,\beta _2 ;}  \\
\end{array}x} \right] = \displaystyle \frac{1}{{\Gamma \left( {\alpha _1 }
\right)}} \displaystyle \int\limits_0^\infty  {} e^{ - \xi } \xi
^{\alpha _1  - 1} {}_2F_2 \left[ {\begin{array}{*{20}c}
{\alpha _2 ,\alpha _3 ;}  \\
{\beta _1 ,\beta _2 ;}  \\
\end{array}x\xi } \right]d\xi ,\,\,{\mathop{\rm Re}\nolimits} \,\alpha _1  > 0,\eqno (4.17)
$$
$$
\begin{array}{l}
 {}_3F_2 \left[ {\begin{array}{*{20}c}
   {\alpha _1 ,\alpha _2 ,\alpha _3 ;}  \\
   {\beta _1 ,\beta _2 ;}  \\
\end{array}x} \right] = \displaystyle \frac{1}{{\Gamma \left( {\alpha _1 }
\right)\Gamma \left( {\alpha _2 } \right)}}\displaystyle \int
\limits_0^\infty  {\displaystyle \int\limits_0^\infty {} } e^{ -
\xi } e^{ - \eta } \xi ^{\alpha _1  - 1} \eta ^{\alpha _2  - 1}
{}_1F_2 \left[ {\begin{array}{*{20}c}
   {\alpha _3 ;}  \\
   {\beta _1 ,\beta _2 ;}  \\
\end{array}x\xi \eta } \right]d\xi d\eta , \\
{\mathop{\rm Re}\nolimits} \,\,\alpha _1  > \,0,\,{\mathop{\rm
Re}\nolimits} \,\,\alpha _2  > 0,
\end{array}\eqno (4.18)
$$
$$
\begin{array}{l}
{}\displaystyle _3F_2 \left[ {\begin{array}{*{20}c}
{\alpha _1 ,\alpha _2 ,\alpha _3 ;}  \\
{\beta _1 ,\beta _2 ;}  \\
\end{array}x} \right] \\
= \displaystyle \frac{1}{{\Gamma \left( {\alpha _1 } \right)\Gamma
\left( {\alpha _2 } \right)\Gamma \left( {\alpha _3 }
\right)}}\displaystyle \int\limits_0^\infty {\int\limits_0^\infty
{\int\limits_0^\infty  {} } } e^{ - \xi  - \eta  - \zeta } \xi
^{\alpha _1  - 1} \eta ^{\alpha _2  - 1} \zeta ^{\alpha _3  - 1}
{}_0F_2 \left[ {\begin{array}{*{20}c} { - ;}  \\
{\beta _1 ,\beta _2 ;}  \\
\end{array}x\xi \eta \zeta } \right]d\xi d\eta d\zeta , \\
{\mathop{\rm Re}\nolimits} \,\,\alpha _i  > 0,\,i = 1,2,3. \\
\end{array}\eqno (4.19)
$$

\section{Formulas of an analytic continuation for Clausen function}

As it was marked in the paper [4] to find the formula of an
analytic continuation it is necessary to reduce the order of
hypergeometric function. For this purpose we use expansion (4.1)
at values $ p = 3,\,\,q = 2$. Then the generalized hypergeometric
function (1.1) looks like, which refers to as Clausen function
([1], p. 141)
$$
{}_3F_2 \left[ {\begin{array}{*{20}c}
{\alpha _1 ,\alpha _2 ,\alpha _3 ;}  \\
{\beta _1 ,\beta _2 ;}  \\
\end{array}z} \right] = \displaystyle \sum\limits_{m = 0}^\infty  {}
\displaystyle \frac{{\left( {\alpha _1 } \right)_m \left( {\alpha
_2 } \right)_m \left( {\alpha _3 } \right)_m }}{{\left( {\beta _1
} \right)_m \left( {\beta _2 } \right)_m m!}}z^m .\eqno (5.1)
$$
\textbf{Theorem}. If for parameters of Clausen function (5.1) it
is satisfied a condition $ \beta _1 ,\beta _2 ,\alpha _2  - \alpha
_1 ,\beta _1  - \alpha _1  - \alpha _2  \ne 0, \pm 1, \pm 2,...,$
then the following formulas of an analytic continuation are fair

$$
\begin{array}{l}
{}_3F_2 \left[ {\begin{array}{*{20}c}
{\alpha _1 ,\alpha _2 ,\alpha _3 ;}  \\
{\beta _1 ,\beta _2 ;}  \\
\end{array}x} \right] \\
= \displaystyle B_1 \left( { - x} \right)^{ - \alpha _1 } F_2
\left( {\alpha _1 ;\beta _2  - \alpha _3 ,1 - \beta _1  + \alpha
_1 ;\beta _2 ,1 - \alpha _2  + \alpha _1 ;1,
\displaystyle \frac{1}{x}} \right) \\
+ B_2 \left( { - x} \right)^{ - \alpha _2 } F_2 \left( {\alpha _2
;\beta _2  - \alpha _3 ,1 - \beta _1  + \alpha _2 ;\beta _2 ,1 - \alpha _1  +
\alpha _2 ;1,\displaystyle \frac{1}{x}} \right), \\
\end{array}\eqno (5.2)
$$
$$
\begin{array}{l}
{}_3F_2 \left[ {\begin{array}{*{20}c}
{\alpha _1 ,\alpha _2 ,\alpha _3 ;}  \\
{\beta _1 ,\beta _2 ;}  \\
\end{array}x} \right] \\
= \displaystyle B_1 \left( {1 - x} \right)^{ - \alpha _1 } F_2
\left( {\alpha _1 ;\beta _2  - \alpha _3 ,\beta _1  - \alpha _2
;\beta _2 ,1 - \alpha _1  - \alpha _2 ;
\displaystyle \frac{x}{{x - 1}},\displaystyle \frac{1}{{1 - x}}} \right) \\
+ \displaystyle B_2 \left( {1 - x} \right)^{ - \alpha _2 }
\displaystyle F_2 \left( {\alpha _2 ;\beta _2  - \alpha _3 ,\beta
_1  - \alpha _1 ;\beta _2 ,1 - \alpha _2  - \alpha _1 ; \displaystyle \frac{x}{{x - 1}},\frac{1}{{1 - x}}} \right), \\
\end{array}\eqno (5.3)
$$
$$
\begin{array}{l}
\displaystyle {}_3F_2 \left[ {\begin{array}{*{20}c}
{\alpha_1 ,\alpha_2 ,\alpha_3 ;}  \\
{\beta_1 ,\beta_2 ;}  \\
\end{array}x} \right]= \displaystyle B_1 F_2
\left( {\alpha_1 ;\alpha_3 ,\beta_1  - \alpha_2 ;\beta_2 ,1 - \alpha_2  + \alpha_1 ;x,1} \right) \\
\,\,\,\,\,\,\,\,\,\,\,\,\,\,\,\,\,\,\,\,\,\,\,\,\,\,\,\,\,\,\,
\,\,\,\,\,\,\,\,\,\,\,\,\,\,\,\,\,\,\,\,\,\,\,\,\,\,+
\displaystyle B_2 \displaystyle F_2 \left( {\alpha_2 ;\alpha_3
,\beta_1  - \alpha_1 ;\beta_2 ,1 -
\alpha_1  + \alpha_2 ;x,1} \right), \\
\end{array}\eqno (5.4)
$$
$$
\begin{array}{l}
\displaystyle {}_3F_2 \left[ {\begin{array}{*{20}c}
   {\alpha _1 ,\alpha _2 ,\alpha _3 ;}  \\
   {\beta _1 ,\beta _2 ;}  \\
\end{array}x} \right] = \displaystyle A_1 F_{1,1,0;}^{2,1,0;} \left[ {\begin{array}{*{20}c}
   {\,\,\,\,\,\,\,\,\,\,\,\,\,\,\,\,\,\,\,\alpha _1 ,\alpha _2 ;}  \\
   {1 + \alpha _1  + \alpha _2  - \beta _1 ;}  \\
\end{array}\begin{array}{*{20}c}
   {\beta _2  - \alpha _3 ;}  \\
   {\,\,\,\,\,\,\,\,\,\,\beta _2 ;}  \\
\end{array}\begin{array}{*{20}c}
   { - ;}  \\
   { - ;}  \\
\end{array}x,1 - x} \right] \\
+ A_2 \left( {1 - x} \right)^{\beta _1  - \alpha _1  - \alpha _2 } H_2
\left( {\alpha _1  + \alpha _2  - \beta _1 ;\beta _2  - \alpha _3 ,\beta _1  -
\alpha _1 ,\beta _1  - \alpha _2 ;\beta _2 ; \displaystyle \frac{x}{{x - 1}},x - 1} \right), \\
\end{array}\eqno (5.5)
$$
where
$$
H_2 \left( {\alpha ;\beta ,\gamma ,\delta ;\varepsilon ;x,y}
\right) = \displaystyle \sum\limits_{m,n = 0}^\infty  {}
\frac{{\left( \alpha \right)_{m - n} \left( \beta  \right)_m
\left( \gamma  \right)_n \left( \delta  \right)_n }}{{\left(
\varepsilon  \right)_m m!n!}}x^m y^n ,
$$
$$
A_1  = \displaystyle \frac{{\Gamma \left( {\beta _1 }
\right)\Gamma \left( {\beta _1  - \alpha _1  - \alpha _2 }
\right)}}{{\Gamma \left( {\beta _1  - \alpha _1 } \right)\Gamma
\left( {\beta _1  - \alpha _2 } \right)}},\,\,\, A_2  =
\frac{{\Gamma \left( {\beta _1 } \right)\Gamma \left( {\alpha _1 +
\alpha _2  - \beta _1 } \right)}}{{\Gamma \left( {\alpha _1 }
\right)\Gamma \left( {\alpha _2 } \right)}},\eqno (5.6)
$$
$$
B_1  = \frac{{\Gamma \left( {\beta _1 } \right)\Gamma \left(
{\alpha _2  - \alpha _1 } \right)}}{{\Gamma \left( {\alpha _2 }
\right)\Gamma \left( {\beta _1  - \alpha _1 } \right)}},\,\,\, B_2
= \frac{{\Gamma \left( {\beta _1 } \right)\Gamma \left( {\alpha _1
- \alpha _2 } \right)}}{{\Gamma \left( {\alpha _1 } \right)\Gamma
\left( {\beta _1  - \alpha _2 } \right)}}.\eqno (5.7)
$$
\textbf{Proof}. Using the formula of an analytic continuation of
Gauss function [2]
$$
\begin{array}{l}
\displaystyle F\left( {a,b;c;x} \right) = \displaystyle \frac{{\Gamma \left( c \right)
\Gamma \left( {b - a} \right)}}{{\Gamma \left( b \right)\Gamma \left( {c - a} \right)}}
\left( { - x} \right)^{ - a} \displaystyle F\left( {a,1 - c + a;1 - b + a;\displaystyle \frac{1}{x}} \right) \\
\,\,\,\,\,\,\,\,\,\,\,\,\,\,\,\,\,\,\,\,\,\,\,\,\,\,\,\,\,\,\,\,\,
+ \displaystyle \frac{{\Gamma \left( c \right) \Gamma \left( {a -
b} \right)}}{{\Gamma \left( a \right)\Gamma \left( {c - b}
\right)}}\left( { - x} \right)^{ - b} \displaystyle F\left( {b,1 - c + b;1 - a + b;\displaystyle \frac{1}{x}} \right), \\
\end{array}\eqno (5.8)
$$
By virtue of the formula of an analytic continuation (5.8) from
expansion (4.1), we have
$$
\begin{array}{l}
{}_3F_2 \left[ {\begin{array}{*{20}c}
{\alpha _1 ,\alpha _2 ,\alpha _3 ;}  \\
{\beta _1 ,\beta _2 ;}  \\
\end{array}x} \right]  \\
= \left( { - x} \right)^{ - \alpha _1 } \displaystyle
\frac{{\Gamma \left( {\beta _1 } \right) \Gamma \left( {\alpha _2
- \alpha _1 } \right)}}{{\Gamma \left( {\alpha _2 } \right) \Gamma
\left( {\beta _1  - \alpha _1 } \right)}} \displaystyle
\sum\limits_{i = 0}^\infty  {}  \displaystyle \frac{{\left(
{\alpha _1 } \right)_i \left( {\alpha _2 } \right)_i \left( {\beta
_2  - \alpha _3 } \right)_i }}{{\left( {\alpha _2 } \right)_i
\left( {\beta _2 } \right)_i i!}} \displaystyle F \left( {\alpha
_1  + i,1 - \beta _1 + \alpha _1 ;1 - \alpha _2  + \alpha _1 ; \displaystyle \frac{1}{x}} \right) \\
+ \left( { - x} \right)^{ - \alpha _2 } \displaystyle
\frac{{\Gamma \left( {\beta _1 } \right) \Gamma \left( {\alpha _1
- \alpha _2 } \right)}}{{\Gamma \left( {\alpha _1 } \right) \Gamma
\left( {\beta _1  - \alpha _2 } \right)}} \displaystyle
\sum\limits_{i = 0}^\infty  {}  \displaystyle \frac{{\left(
{\alpha _1 } \right)_i \left( {\alpha _2 } \right)_i \left( {\beta
_2  - \alpha _3 } \right)_i }}{{\left( {\alpha _1 } \right)_i
\left( {\beta _2 } \right)_i i!}}F\left( {\alpha _2  + i,1 - \beta
_1  + \alpha _2 ;1 - \alpha _1  + \alpha _2 ; \displaystyle \frac{1}{x}} \right). \\
\end{array}\eqno (5.9)
$$
Decomposing in a series hypergeometric of Gauss function in
identity (5.9) and considering definition of Appell function $ F_2
\left( {a;b_1 ,b_2 ;c_1 ,c_2 ;x,y} \right)$, we receive the
formula analytic continuation (5.2) for Clausen function. It is
similarly proved also other formulas (5.3) - (5.5). The theorem is
proved.

\end{document}